\documentclass[aps,superscriptaddress,showpacs]{revtex4}
\usepackage{amssymb}
\usepackage{amsmath}
\usepackage{graphicx}
\usepackage[normalem]{ulem}

\usepackage[dvips]{color}

\def\esym{$E_{sym}(\rho)$~}

\def\es0{$E_{sym}(\rho_0)$~}

\def\usym{$U_{sym}(\rho,k)$~}

\def\us0{$U_{sym}(\rho_0,k_F)$~}

\def\lr{$L(\rho)$~}

\def\l0{$L(\rho_0)$~}

\def\emass{$m^*_{n-p}(\rho_0,\delta)$~}

\renewcommand\sout{\bgroup \color{red} \ULdepth=-.5ex \ULset}

\begin{document}
\title{Constraining the neutron-proton effective mass splitting using empirical constraints on the density dependence of nuclear symmetry energy around normal density}
\author{Bao-An Li}\email{Bao-An.Li@tamuc.edu}
\affiliation{Department of Physics and Astronomy, Texas A$\&$M
University-Commerce, Commerce, TX 75429-3011, USA}
\affiliation{Department of Applied Physics, Xi'an Jiaotong University, Xi'an 710049, China}
\author{Xiao Han}
\affiliation{Department of Applied Physics, Xi'an Jiaotong University, Xi'an 710049, China}

\begin{abstract}
According to the Hugenholtz-Van Hove theorem, nuclear symmetry energy \esym and its slope \lr at an arbitrary density $\rho$ are determined by the nucleon isovector (symmetry) potential \usym and its momentum dependence $\frac{\partial U_{sym}}{\partial k}$. The latter determines uniquely the neutron-proton effective k-mass splitting $m^*_{n-p}(\rho,\delta)\equiv (m_{\rm n}^*-m_{\rm p}^*)/m$ in neutron-rich nucleonic matter of isospin asymmetry $\delta$. Using currently available constraints on the \es0 and \l0 at normal density $\rho_0$ of nuclear matter from 28 recent analyses of various terrestrial nuclear laboratory experiments and astrophysical observations, we try to infer the corresponding neutron-proton effective k-mass splitting $m^*_{n-p}(\rho_0,\delta)$. While the mean values of the $m^*_{n-p}(\rho_0,\delta)$ obtained from most of the studies are remarkably consistent with each other and scatter very closely around an empirical value of \emass$=0.27\cdot\delta$, it is currently not possible to scientifically state surely that the \emass is positive within the present knowledge of the uncertainties. Quantifying, better understanding and then further reducing the uncertainties using modern statistical and computational techniques in extracting the \es0 and \l0 from analyzing the experimental data are much needed.
\end{abstract}
\maketitle

\section{Introduction}
The ultimate goal of investigating properties of neutron-rich nucleonic matter through terrestrial nuclear laboratory experiments and
astrophysical observations is to understand the underlying isospin dependence of strong interaction in nuclear medium \cite{LRP}.
The Equation of State (EOS) of neutron-rich nucleonic matter can be written within the parabolic approximation in terms of the binding
energy per nucleon at density $\rho$ as $E(\rho ,\delta)=E(\rho ,\delta =0)+E_{\rm sym}(\rho )\delta^{2}+\mathcal{O}(\delta^4)$
where $\delta\equiv(\rho_{n}-\rho _{p})/(\rho _{p}+\rho _{n})$ is the neutron-proton asymmetry and $E_{\rm sym}(\rho)$ is the
density-dependent nuclear symmetry energy. The latter has important applications in many areas of both nuclear physics,
see, e.g., refs. \cite{ireview98,ibook01,dan02,ditoro,LCK08,Lynch09,Trau12} and astrophysics, see, e.g., refs. \cite{lat01,Steiner05,Lattimer12}.
However, the density dependence of nuclear symmetry energy has been among the most uncertain properties of neutron-rich nucleonic matter. Predictions using various many-body theories and interactions diverge quite broadly especially at abnormal densities. It is thus exciting to see that significant progress has been made recently in constraining the \esym around $\rho_0$, see, e.g., ref.\ \cite{Tsang12} based on model analyses of experimental and/or observational data. In particular, as listed in Table 1 and also shown in Fig. 1 at least 28 studies have extracted the slope $L(\rho_0) \equiv \left[3 \rho (\partial E_{\rm sym}/\partial \rho\right]_{\rho_0}$ and $E_{\rm sym}(\rho_0)$ at $\rho_0$ \cite{LWChen11,Moller,agr12,Pawel13,sun10,tsa1,tsa2,chen05a,li05a,shet,Jim,Cen09,War09,Chen10,Mliu10,XuLiChen10a,Kli07,Car10,mye96,Koh10,Dong13a,Dong13b,Zhang13,Tamii13,Vid12,Wen12,Steiner10,Steiner12,Mike,Sot12,Newton09}.
It is thus interesting to ask timely what we can learn about the isospin dependence of in-medium nuclear interaction from the extracted constrains on \l0 and \es0. Here we study this question at the mean-field level by using a formulism developed earlier in Refs. \cite{XuLiChen10a,xuli2,Rchen} based on the Hugenholtz-Van Hove (HVH) theorem \cite{hug}. Specifically, we try to infer both the magnitude of the symmetry potential $U_{\rm sym}(\rho_0,k_F)$ and the neutron-proton effective k-mass splitting $m^*_{n-p}(\rho_0,\delta)$ corresponding to each of the 28 constraints on \es0 and \l0 at $\rho_0$. The consistency of the extracted values for $U_{\rm sym}(\rho_0,k_F)$ and $m^*_{n-p}(\rho_0,\delta)$ from various constraints is then examined. It is found that while the mean values of the $U_{\rm sym}(\rho_0,k_F)$ and $m^*_{n-p}(\rho_0,\delta)$ from different studies are consistent with each other and most of them scatter closely around \us0=29 MeV and \emass$=0.27\cdot\delta$, respectively, the individual uncertainties from many analyses are still too large. Quantifying, better understanding and reducing the uncertainties in extracting the symmetry energy from  model analyses of the experimental data are much needed in order to use reliably the extracted mean values of the $U_{\rm sym}(\rho_0,k_F)$ and $m^*_{n-p}(\rho_0,\delta)$ in solving many important problems in both nuclear physics and astrophysics.
\begin{table}[tbp]
\caption{{\protect\small Constrained values of \es0 and \l0 from 28 analyses of terrestrial nuclear experiments and astrophysical observations}}
\label{data}%
\begin{tabular}{ccccccc}
\hline\hline
Analysis & \es0 & \l0 & Ref. \\ \hline
Thomas-Fermi model analysis of masses (Myers 1996) & $32.65$ & $50$ &\cite{mye96}\\
Atomic masses (Liu 2010) & $31.1\pm 1.7$& $66\pm 13$ & \cite{Mliu10}\\
Liquid drop model analysis of atomic masses (Lattimer 2012) & $29.6\pm 3.$& $46.6\pm 37$ & \cite{Jim}\\
FRDM analysis of atomic masses (Moller 2012)& $32.5\pm 0.5$ & $70\pm 15$ & \cite{Moller}\\
Atomic masses and n-skin of Sn isotopes (Chen 2011)& $30.5\pm 3$ & $52.5\pm 20$ & \cite{LWChen11}\\
Atomic masses and n-skin in an empirical approach (Agrawal 2012)& $32.1$ & $64\pm 5$ & \cite{agr12}\\
IAS+n-skin (Danielewicz and Lee 2013)& $31.95\pm 1.75$ & $52.5\pm 17.5$ & \cite{Pawel13}\\
SHF+n-skin (Chen 2010) & $30.5\pm 5.5$& $41\pm 41$ & \cite{Chen10}\\
Droplet Model+n-skin (Centelles \& Warda 2009) & $31.5\pm 3.5$& $55\pm 25$ & \cite{Cen09,War09}\\
IBUU04 analysis of isospin diffusion at 50 MeV/A (Chen \& Li 2005) & $31.6$& $86\pm 25$ & \cite{chen05a,li05a}\\
IQMD analysis of isospin diffusion at 50 MeV/A (Tsang 2009) & $32.5\pm 2.5$& $77.5\pm 32.5$ & \cite{tsa1,tsa2}\\
IQMD analysis of isospin diffusion at 35 MeV/A (Sun 2010) & $30.1$& $52$ & \cite{sun10}\\
Isoscaling analysis of fragments (Shetty 2007) & $31.6$& $65$ & \cite{shet}\\
Global nucleon optical potential (Xu 2010) & $31.3\pm 4.5$& $52.7\pm 22.5$ &\cite{XuLiChen10a}\\
Pygmy dipole resonances (Klimkiewicz 2007) & $32\pm 1.8$ & $43 \pm 15 $ &\cite{Kli07}\\
Pygmy dipole resonances (Carbone 2010) & $32\pm 1.3$ & $65 \pm 16 $ &\cite{Car10}\\
AMD analysis of transverse flow (Kohley 2010) & $30.5$ & $65$ &\cite{Koh10}\\
$\alpha$-decay energy (Dong 2013) & $31.6\pm 2.2$ & $61\pm 22$ &\cite{Dong13a}\\
$\beta$-decay energy (Dong 2013) & $32.3\pm 1.3$ & $50\pm 15$ &\cite{Dong13b}\\
Mass differences and n-skin (Zhang 2013) & $32.3\pm 1.0$ & $45.2\pm 10$ &\cite{Zhang13}\\
Dipole polarizability of $^{208}$Pb (Tamii 2013) & $30.9\pm 1.5$ & $46\pm 15$ &\cite{Tamii13}\\
r-mode instability of neutron stars (Vidana 2012) & $30.\pm 5$ & $\geq 50$ &\cite{Vid12}\\
r-mode instability of neutron stars (Wen 2012) & $32.5\pm 7.5$ & $\leq 65$ &\cite{Wen12}\\
Mass-radius of neutron stars-analysis1 (Steiner 2010) & $31\pm 3$ & $50\pm 10$ &\cite{Steiner10}\\
Mass-radius of neutron stars-analysis2 (Steiner 2012) & $33\pm 1.6$ & $46\pm 10$ &\cite{Steiner12}\\
Torsional crust oscillation of neutron stars (Gearheart 2011) & $32.5\pm 7.5$ & $\leq 50 $ &\cite{Mike}\\
Torsional crust oscillation of neutron stars (Sotani 2012) & $32.5\pm 7.5$ & $115\pm 15$ &\cite{Sot12}\\
Binding energy of neutron stars (Newton 2009) & $32.5\pm 7.5$ & $\leq 70 $ &\cite{Newton09}\\
\hline\hline
\end{tabular}%
\end{table}
\begin{figure}[htb]
\begin{center}
\includegraphics[width=17cm,height=7cm]{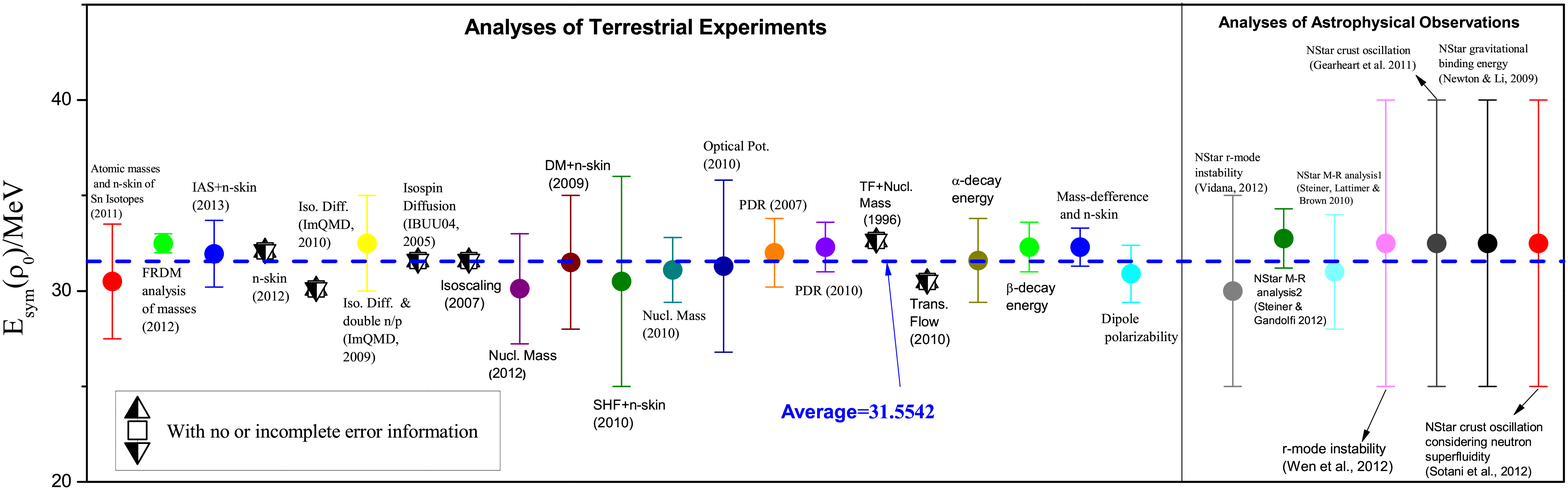}\label{E-Lsym}
\includegraphics[width=17cm,height=7cm]{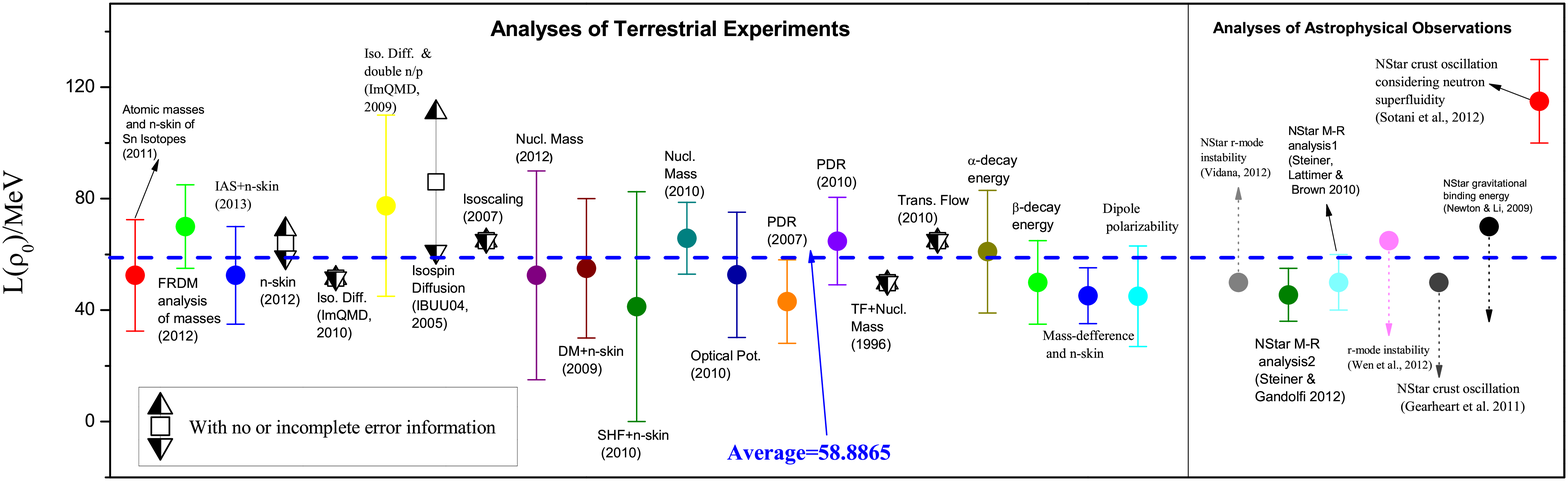}
\caption{(Color online) Nuclear symmetry energy (upper) and its slope L (lower) at normal density of nuclear matter from 28 analyses of terrestrial nuclear laboratory experiments and astrophysical observations.}
\end{center}
\end{figure}
\section{Relationship between neutron-proton effective mass splitting and symmetry energy based on the Hugenholtz-Van Hove theorem}
According to the well-known Lane potential \cite{Lan62} verified by various many-body theories and optical model analysis of nucleon-nucleus scattering data, the neutron/proton (n/p) single-particle potential $U_{n/p}(\rho,k,\delta)$ can be well approximated by
\begin{equation}\label{Lane0}
U_{\rm n/ \rm p}(\rho,k,\delta)=U_0(\rho,k) \pm U_{\rm sym}(\rho,k)\cdot\delta+\mathcal{O}(\delta^2),
\end{equation}
where the $U_0(\rho,k)$ and $U_{\rm sym}(\rho,k)$ are, respectively, the nucleon isoscalar and isovector (symmetry) potentials for nucleons with momentum $k$ in asymmetric nuclear matter of isospin asymmetry $\delta$ at density $\rho$. Their momentum dependence is normally characterized by the nucleon effective k-mass
\begin{equation}
m_{\tau}^*/m=[1+\frac{m}{\hbar^2 k_F} \frac{d U_{\tau}}{d k}|_{k_F} ]^{-1}
\end{equation}
where $\tau=$n, p and 0 for neutrons, protons and nucleons, respectively, and $m=(m_n+m_p)/2$ is the average mass of nucleons in free-space. While the nucleon isoscalar potential and its momentum dependence, especially at $\rho_0$, have been relatively well determined, our knowledge about the isovector potential \usym and its momentum dependence $\frac{\partial U_{sym}}{\partial k}$ even at normal density is still very poor. However, from the structure of rare isotopes and mechanism of heavy-ion reactions to the cooling of protoneutron stars, solutions to many interesting issues depend critically on the nucleon isovector potential and its momentum dependence.

Using the Brueckner theory \cite{bru64} or the Hugenholtz-Van Hove (HVH) theorem \cite{hug}, the \esym and \lr can be expressed as \cite{XuLiChen10a,xuli2,Rchen,Dab73}
\begin{eqnarray}
&& E_{\rm sym}(\rho) = \frac{1}{3} \frac{\hbar^2 k_F^2}{2 m_0^*} +
\frac{1}{2} U_{\rm sym}(\rho,k_{F}), \label{Esymexp2}
\\
&& L(\rho) = \frac{2}{3} \frac{\hbar^2 k_F^2}{2 m_0^*} +
\frac{3}{2} U_{\rm sym}(\rho,k_F) + \frac{\partial U_{\rm sym}}{\partial k}|_{k_F} k_F, \label{Lexp2}
\end{eqnarray}
where $k_F=(3\pi^2\rho/2)^{1/3}$ is the nucleon Fermi momentum. We emphasize that these relationships are general and independent of the many-body theory and/or interaction used to calculate the \usym and $m^*_0$. In fact, all microscopic calculations of the nuclear EOS are required to satisfy the HVH theorem. It is also worth noting that adding the second-order symmetry potential $U_{sym,2}(\rho,k)\cdot\delta^2$ term to the Lane potential in Eq. \ref{Lane0} and considering the $\delta^2$ terms consistently in applying the HVH theorem, while the expression for the $E_{\rm sym}(\rho)$ remains the same as in Eq. \ref{Esymexp2}, the expression for $L(\rho)$ has two additional terms due to the momentum dependence of the isoscarlar effective mass $m_0^*$ and the $U_{sym,2}(\rho,k_F)$, respectively \cite{Rchen}. However, at the saturation density $\rho_0$ these high-order terms were found completely negligible based on the optical model analyses of the latest and most complete neutron-nucleus scattering data base \cite{xhli13}. Thus, at least at $\rho_0$ the Eqs. \ref{Esymexp2} and \ref{Lexp2} are accurate decompositions of the symmetry energy and its density slope required by the HVH theorem.  While it is not clear if all models satisfy the HVH theorem and the resulting equations \ref{Esymexp2} and \ref{Lexp2}, it is understandable that various observables may be sensitive to different
components of the \esym and \lr with different sensitivities, leading to the rather broad ranges of uncertainties and/or error bars in the results shown in Table 1 and Fig.1. It is certainly an interesting task to find out for each observable whether/why it may only depend on the total values or some particular components of the \esym and/or \lr. We notice that
not all models used in extracting the \es0 and \l0 consider all the terms of the \esym and \lr in the equations \ref{Esymexp2} and \ref{Lexp2}.  For instance, while most models consider the momentum dependence of the isoscalar potential albeit often use different values for the $m^*_0$, the momentum dependence of the isovector potential, i.e., the $ \frac{\partial U_{\rm sym}}{\partial k}$ term, has been frequently ignored so far. It may well be that some of the observables are not sensitive to this component of the \lr but still allow an accurate extraction of the \es0 and \l0 within the framework of a given model used. In this work, we use the 28 sets of \es0 and \l0 as quasi-data regardless how they were extracted from the model analyses of experimental data. Since the expressions for \esym and \lr in equations \ref{Esymexp2} and \ref{Lexp2} are generally required by the HVH theorem, using the most widely used empirical value of  $m^*_0$, we can infer from the quasi-data the required values of the neutron-proton effective mass splitting to satisfy the  equations \ref{Esymexp2} and \ref{Lexp2}. Whether such an effective mass splitting is consistently predicted in each model used is an interesting question worth a careful study.

Since the $m^*_0$ is well determined at $\rho_0$, given the values of $E_{\rm sym}(\rho_0)$ and $L(\rho_0)$, the $U_{\rm sym}(\rho_0,k_F)$ and $\frac{\partial U_{\rm sym}}{\partial k}|_{k_F}$ are then uniquely determined by the
equations \ref{Esymexp2} and \ref{Lexp2}. We stress here that the HVH theorem requires the $U_{\rm sym}(\rho_0,k_F)$ and $\frac{\partial U_{\rm sym}}{\partial k}|_{k_F}$ (or equivalently the \es0 and \l0) to be correlated as they are both determined by the same energy density functional \cite{xuli2,Rchen,xu00}. Thus, they should not be independently varied. More explicitly, a simple inversion leads to $U_{\rm sym}(\rho_0,k_F) = 2[E_{\rm sym}(\rho_0)-\frac{1}{3}\frac{m}{m^*_0}E_{F}(\rho_0)]$ and $(\frac{dU_{\rm sym}}{dk})_{k_{F}}(\rho_0) =[L(\rho_0)-3E_{\rm sym}(\rho_0)+\frac{1}{3}\frac{m}{m^*_0}E_{F}(\rho_0)]/k_F$ where $E_F(\rho_0)$ is the Fermi energy at $\rho_0$. It is seen that while the \us0 is completely determined by the \es0 and $m/m^*_0$, the $(\frac{dU_{\rm sym}}{dk})_{k_{F}}(\rho_0)$ also depends on the \l0. It is well known that for a given set of two-body and three-body nuclear interactions, the resulting nucleon potential often depends on the many-body theory used. On the other hand, the single-particle mean-field potential is often the one directly tested in comparing model calculations with experimental/observational data. For example, it is the input for most shell model calculations of nuclear structure and transport model simulations of nuclear reactions. The expressions
\ref{Esymexp2} and \ref{Lexp2} for \esym and \lr indicate that one can use the density dependence of nuclear symmetry energy extracted from experiments/observations to test directly the nuclear isovector potential and its momentum dependence, or vice versa, without the hinderance of remaining difficulties and uncertainties in nuclear many-body theories. Here, we are interested in learning about the isospin dependence of in-medium nuclear interaction at $\rho_0$ from the constrained \es0 and \l0.

The nucleon effective mass describes to leading order effects related to the non-locality of the underlying nuclear interactions and the Pauli exchange effects in many-fermion systems \cite{mahaux,jamo,dob}. While the nucleon isoscalar effective k-mass is well determined to be $m_0^*/m=0.7\pm0.05$ at $\rho_0$ \cite{jamo}, essentially nothing is known about the nucleon isovector effective mass \cite{dob}. Knowledge about the neutron-proton effective mass splitting is essential for understanding many interesting questions in both nuclear physics and astrophysics \cite{Steiner05,sjo76,neg81,cop85,bethe,pan92,IBUU04}, such as, pairing and superfluidity in nuclei and neutron stars, properties of rare isotopes, isospin transport in heavy-ion reactions, thermal and transport properties of neutron star crust and cooling mechanism of protoneutron stars. Unfortunately, even the sign of the neutron-proton effective mass splitting, not to mention its magnitude, has been a longstanding and controversial issue. While some theories predict that $m_{\rm n}^*\geq m_{\rm p}^*$, the opposite has often been shown by studies using different models or interactions, see, e.g., refs. \cite{ditoro,LCK08,mahaux,jamo,dob,far01,Bali04,stone,zuo05,Fuchs1,Fuchs2,Les06,Ouli}. Thus, a convincing conclusion on this issue will have profound ramifications in both nuclear physics and astrophysics. The momentum dependence of the isovector potential is conventionally measured by using the neutron-proton effective mass splitting
\begin{equation}\label{npemass1}
m^*_{n-p}(\rho_0,\delta)\equiv\frac{m_{\rm n}^*-m_{\rm p}^*}{m}=\frac{\frac{m}{\hbar^2k_{\rm F}}(dU_p/dk-dU_n/dk)}{(1+\frac{m}{\hbar^2k_{\rm F}}dU_p/dk)(1+\frac{m}{\hbar^2k_{\rm F}}dU_n/dk)}\bigg|_{k_{\rm F}}.
\end{equation}
In the above expression, the numerator $\frac{m}{\hbar^2k_{\rm F}}(dU_p/dk-dU_n/dk)$ is exactly $-2 \delta \frac{m}{\hbar^2k_{\rm F}}\frac{dU_{\rm sym}}{dk}$ according to the Lane potential in Eq. \ref{Lane0}. Since the $U_{sym}(\rho,k)\cdot\delta$ term is always much smaller than the isoscalar potential $U_0(\rho,k)$, see, e.g., ref. \cite{kho96}, the denominator can be well approximated by $(1+\frac{m}{\hbar^2k_{\rm F}}dU_p/dk)(1+\frac{m}{\hbar^2k_{\rm F}}dU_n/dk)\approx (1+\frac{m}{\hbar^2k_{\rm F}}dU_0/dk)^2=(m/m^*_0)^2$. We note here that this approximation is slightly different from that used earlier in \cite{XuLiChen10a}. In the latter, an unnecessary approximation $(1+\frac{m}{\hbar^2k_{\rm F}}dU_0/dk)^2\approx(1+2\frac{m}{\hbar^2k_{\rm F}}dU_0/dk)=2\frac{m}{m^*_0}-1$ which is good for $m^*_0\approx m$ was used. Inserting the expression for $(\frac{dU_{\rm sym}}{dk})_{k_{F}}(\rho_0)$ in terms of \es0 and \l0, we then have
\begin{equation}
m^*_{n-p}(\rho_0,\delta)\approx\delta\cdot \left[3E_{\rm sym}(\rho_0)-L(\rho_0)-\frac{1}{3}\frac{m}{m^*_0}E_{F}(\rho_0)\right]\bigg/ \left[E_F(\rho_0)\cdot (m/m_0^*)^2\right]. \label{npemass2}
\end{equation}
It is clear that whether the $m_{\rm n}^*$ is equal, larger or smaller than the $m_{\rm p}^*$ depends on the value of $L(\rho_0)$ relative to the quantity $[3E_{\rm sym}(\rho_0)-\frac{1}{3}\frac{m}{m^*_0}E_{F}(\rho_0)]$. For example, using the most widely accepted empirical vales of \es0=31 MeV, $m_0^*/m=0.7$ and $E_F(\rho_0)=36$ MeV,
to obtain a $m^*_{n-p}(\rho_0,\delta)\geq 0$ a value of \l0$\leq 76$ MeV is required.
\begin{figure}[htb]
\begin{center}
\includegraphics[width=17cm,height=7cm]{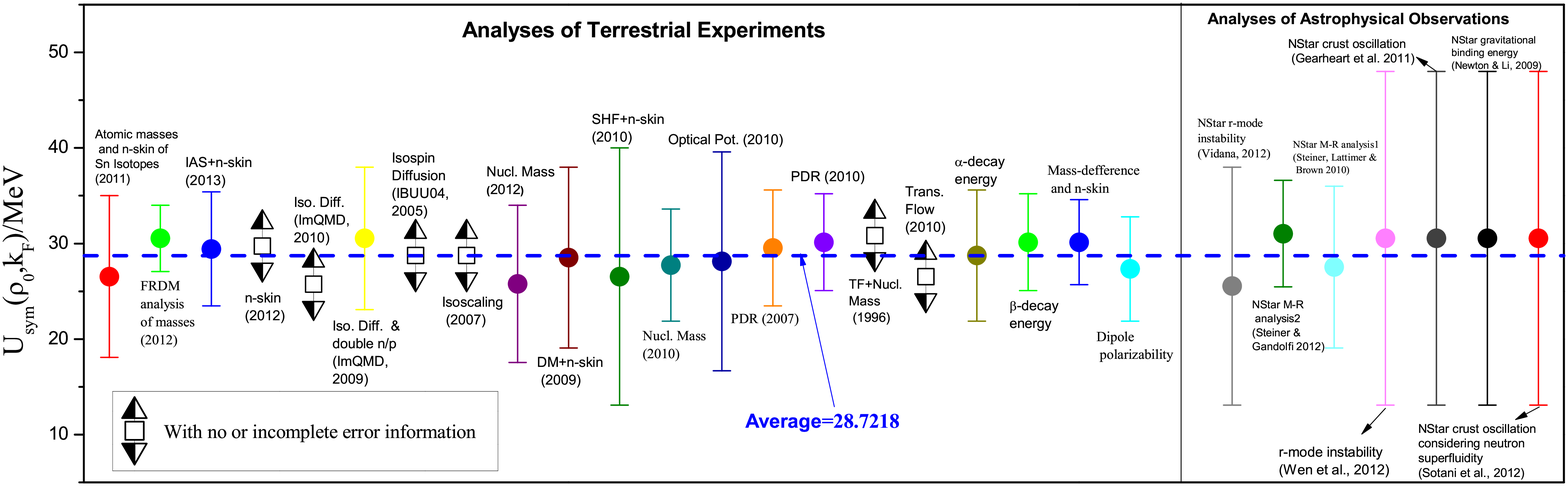}\label{Usym}
\includegraphics[width=17cm,height=7cm]{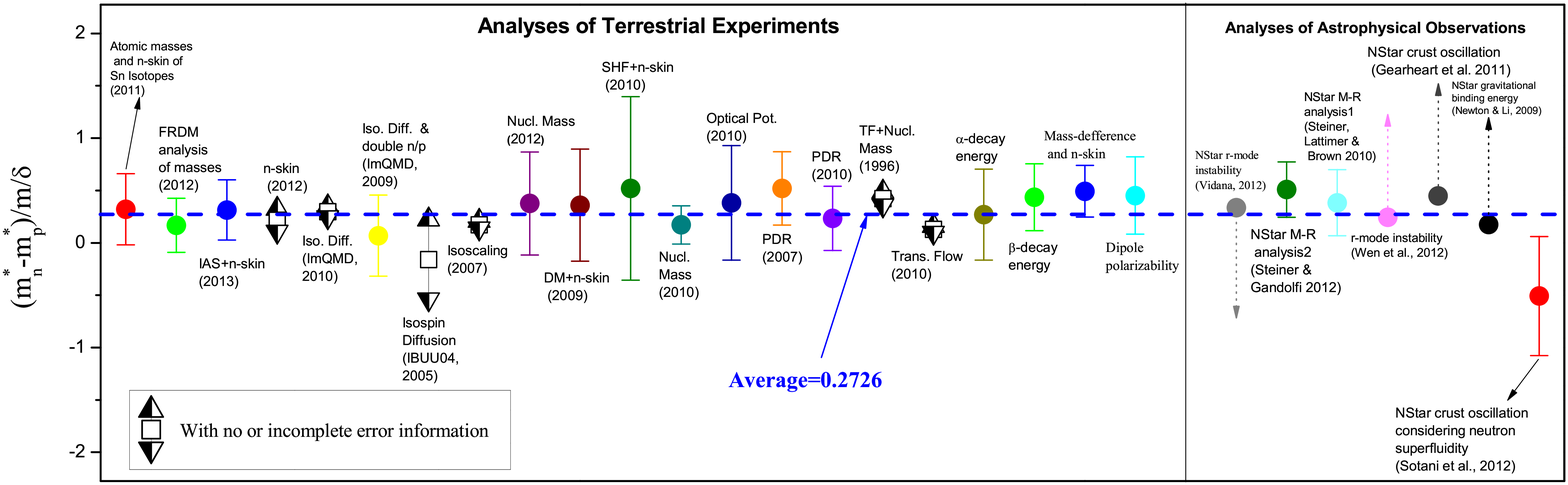}
\caption{(Color online) Nucleon isovector potential \us0 (upper) and neutron-proton effective mass splitting $m^*_{n-p}(\rho_0,\delta)/\delta$ (lower) at normal density of nuclear matter from 28 analyses of terrestrial nuclear laboratory experiments and astrophysical observations.}
\end{center}
\end{figure}
\section{Neutron-proton effective mass splitting from constrains on the density dependence of nuclear symmetry energy around normal density}
It is known that essentially all proposed nuclear interactions have been used in various many-body theories to predict the \esym \cite{LCK08}. Instead of using pure model prediction, we use here the \es0 and \l0 extracted from analyzing terrestrial nuclear laboratory experiments and astrophysical observations. Naturally, all analyses are based on some models and often different approaches are used in analyzing the same data or observations. For instance, at least 5 different models have been used to extract independently the \es0 and \l0 from studying atomic masses. Remarkably, however, with very few exceptions, constraints on the \es0 and \l0 from various analyses of the same or different experiments/observations overlaps closely. Listed in the Table 1 are 28 sets of constraints including the 4 astrophysical ones where only the upper or lower limit of \l0 is given. We notice here that while some of the reported constraints provide both the upper and lower limits or the standard deviation together with the mean values, some do not provide any information about the associated uncertainties but only the mean values of \es0 and \l0. This drawback will be carried over into calculating the corresponding $U_{\rm sym}(\rho_0,k_F)$ and $m^*_{n-p}(\rho_0,\delta)$. It is worth noting here that some of the uncertainties are due to the dual or multiple sensitivities of the selected experimental observables to the symmetry energy and other uncertain ingredients in the model used for the data analyses. For example, the range for \l0 from the IBUU04 transport model analysis \cite{chen05a,li05a} of the isospin diffusion data \cite{tsa1} is mainly due to the undetermined isospin dependence of the in-medium nucleon-nucleon cross sections. The model used a constant \es0=31.6 MeV but adjusted the value of \l0 as well as the in-medium nucleon-nucleon cross sections. The \l0 is equally probable within the range extracted from the analysis. Shown in Fig. 2 are the nucleon isovector potential (upper window) and neutron-proton effective mass splitting (lower window) at $\rho_0$ from the 28 constraints. We caution here that in cases where no or incomplete information about error bars or ranges for the \es0 and/or \l0 was given, only the error bar of the empirical value of $m_0^*/m$ is used in estimating the upper and lower limits of $U_{\rm sym}(\rho_0,k_F)$ and $m^*_{n-p}(\rho_0,\delta)$. These results shown with the black up-down arrows do not have the proper error bars or ranges. In several astrophysical cases where only the upper or lower limits of \l0 were given, the limiting values were indicated with arrows for comparisons. We also must notice that in some cases where correlations between the \es0 and \l0 are considered in certain constrained areas or contours, the maximum ranges are used for both \es0 and \l0 and they are then assumed to be independent. Ideally, the correlations should be maintained. However, most of the available constraints on \es0 and \l0 do not provide any information about such correlation. Nevertheless, it is interesting to see that despite of the large uncertainty ranges of some of the constraints on \es0 and \l0, the resulting mean values of $U_{\rm sym}(\rho_0,k_F)$ and $m^*_{n-p}(\rho_0,\delta)$ from different studies scatter very closely around their global averages of \us0=29 MeV and \emass$=0.27\cdot\delta$, respectively, indicating a high level of consistency of different studies. Moreover, the majority of the inferred $m^*_{n-p}(\rho_0,\delta)$ are positive.
While the mean values of $U_{\rm sym}(\rho_0,k_F)$ and $m^*_{n-p}(\rho_0,\delta)$ are useful in their own rights, to use them reliably as a useful reference for calibrating nuclear many-body theories and much needed inputs for investigating many interesting issues in both nuclear physics and astrophysics, the community should strive at quantifying, better understanding and then further reducing the uncertainties using modern statistical and computational techniques in extracting the \es0 and \l0 from the experimental data. In this regard, it is encouraging to note that some concerted efforts in this direction are under way.

\section{Conclusion}
Based on the Hugenholtz-Van Hove theorem, nuclear symmetry energy and the neutron-proton effective k-mass splitting are explicitly related to each other. Available constraints on the symmetry energy can be used to infer directly the poorly known but very important neutron-proton effective k-mass splitting in neutron-rich nucleonic matter. As an example, we have shown that the constraints on nuclear symmetry energy \es0 and its density slope \l0 at $\rho_0$ from 28 studies of terrestrial nuclear laboratory experiments and astrophysical observations indicate consistently that the nuclear isovector potential and neutron-proton effective k-mass splitting at $\rho_0$ are approximately \us0=29 MeV and \emass$=0.27\cdot\delta$, respectively. Because some constraints on the \es0 and \l0 are given in certain ranges in which all values are equally probable, some others are given in terms of the means and the standard deviations, while the rest are given with only the mean values without any information about the associated uncertainties, we find it is currently impossible to give a physically meaningful ``error bar" for the global averages assuming all reported constraints are statistically independent. As a reference for future comparisons and to illustrate further the importance of quantifying, better understanding and reducing the uncertainties, the current global averages, ``standard deviations" obtained using the 28 mean values and the average sizes of the uncertainty ranges when available are summarized for the \es0,\l0, \us0 and \emass in Table 2.
\begin{table}[tbp]
\caption{{\protect\small 2013 global averages, ``standard deviations" and average sizes of ``error bars" of \es0, \l0, \us0 and \emass from 28 analyses available}}
\label{data}%
\begin{tabular}{ccccccc}
\hline\hline
Quantity: & \es0 (MeV) & \l0 (MeV) & \us0 (MeV) &\emass $(\delta)$ \\ \hline
2013 global average & $31.6$ & $58.9$ & 28.7 & 0.27\\
``Standard deviation" & $0.92$ & $16.5$ & 1.82 & 0.25\\
Average of ``error bars" & $2.66$ & $16.0$ & 7.78 & 0.35\\
\hline\hline
\end{tabular}%
\end{table}
While the mean values from most analyses are rather consistent and point toward a positive \emass, it is currently not possible to scientifically state surely that the \emass is positive within the present knowledge of the uncertainties.

\section{Acknowledgement}
We would like to thank Lie-Wen Chen, Farrooh Fattoyev, William Newton and Chang Xu for helpful discussions. This work is supported in part by the
National Science Foundation under Grant No. PHY-1068022, National Aeronautics and Space Administration under grant NNX11AC41G issued through the Science Mission Directorate and the CUSTIPEN (China-U.S. Theory Institute for Physics with Exotic Nuclei) under DOE grant number DE-FG02-13ER42025.

\end{document}